Quantitative imaging of hybrid chiral spin textures in magnetic multilayer systems by Lorentz microscopy

K. Fallon[1], S. McVitie[1], W. Legrand[2], F. Ajejas[2], D. Maccariello[2], S. Collin[2], V. Cros[2], N. Reyren[2]

[1]*Scottish Universities Physics Alliance, School of Physics and Astronomy, University of Glasgow, Glasgow, G12 8QQ, United Kingdom*

[2]*Unité Mixte de Physique, CNRS, Thales, Univ. Paris-Sud, Université Paris-Saclay, 91767, Palaiseau, France*

Abstract

Chiral magnetic textures in ultrathin perpendicularly magnetised multilayer film stacks with an interfacial Dzyaloshinskii-Moriya interaction have been the focus of much research recently. The chirality associated with the broken inversion symmetry at the interface between an ultrathin ferromagnetic layer and a heavy metal with large spin-orbit coupling supports homochiral Néel domain walls and hedgehog (Néel) skyrmions. Under spin-orbit torques these Néel type magnetic structures are predicted, and have been measured, to move at high velocities. However recent studies have indicated that some multilayered systems may possess a more complex hybrid domain wall configuration, due to the competition between interfacial DMI and interlayer dipolar fields. These twisted textures are expected to have thickness dependent Néel and Bloch contributions to the domain or skyrmion walls. In this work, we use the methods of Lorentz microscopy to measure quantitatively for the first time experimentally both; i) the contributions of the Néel and Bloch contributions and ii) their spatial spin variation at high resolution. These are compared with modelled and simulated structures which are in excellent agreement with our experimental results. Our quantitative analysis provides powerful direct evidence of the Bloch wall component which exists in these hybrid walls and will be significant when exploiting such phenomena in spintronic applications.

The Dzyaloshinskii-Moriya interaction (DMI) has proved to be of great interest in the study of magnetic materials whereby an antisymmetric exchange interaction causes rotation of neighbouring spins [1,2]. In magnetic thin film systems, this effect is prominent due to the strong spin-orbit coupling arising, in most cases, from the interfacial exchange interaction between the magnetic moments and neighbouring heavy metal atoms in a multi-layer system [3,4]. The rotation of spins is a result of the exchange Hamiltonian due to the DMI, given by $H = \boldsymbol{D}_{ij} \cdot (\boldsymbol{S}_i \times \boldsymbol{S}_j)$ where $\boldsymbol{D}_{ij}$ is the DMI vector and $\boldsymbol{S}_i$ and $\boldsymbol{S}_j$ are neighbouring spin vectors [5]. The interfacial DMI influences the chiral texture of domain walls in such systems so that a chiral Néel wall configuration is energetically favoured over the usual achiral divergence free Bloch rotation of the magnetisation expected in thin films [6,7]. The handedness of this Néel twisting of spins depends on the non-magnetic material (which determines the direction of $\boldsymbol{D}$) as well as its position with respect to the magnetic layer, i.e. above or below [8,9]. In thin film systems layer combinations such as Pt|Co|Ir have been used to demonstrate chiral magnetic textures with a $|\boldsymbol{D}|$ value up to 2 mJ/m$^2$ [10]. By creating multilayer repeats of these structures, for specific conditions, homochiral walls can be obtained and skyrmions can be stabilised in all layers – due to the combination of DMI and the influence of dipolar interaction in such PMA systems. Indeed, these magnetic/metallic layer systems are likely to have potential applications in future spintronic based data storage and logic devices which would exploit the efficient manipulation of Néel type domain walls and skyrmions [11-16]. However, recent studies have found the existence of hybrid domain walls in multilayer systems in which the interlayer interactions overcome the DMI [17-19]. This type of wall has a three-dimensional structure where, in some cases, internal magnetic layer(s) have a Bloch wall configuration which is situated between Néel wall surface layer(s) with opposite chirality and results in a flux closure configuration through the thickness of the multilayer. This hybrid structure has been inferred by simulations and imaging of the sense of rotation (or effective chirality) in the top layers through circular dichroism (CD) in x-ray resonant magnetic scattering (XRMS) [17]. This study also demonstrated, through simulation, the profound effect that this hybrid structure may have on the motion of walls and/or skyrmions under spin polarised currents; this is a conclusion also reached by a recent theoretical study which also outlines an analytical model for the hybrid structure [18]. In a separate study, simulations and nitrogen vacancy centre spin reconstruction were used to probe the wall structure of a multilayer from measurement of its stray field [19]. Our aim in this paper is to directly image the Bloch structure of the hybrid wall for the first time using Lorentz TEM and to quantify the contributions for each wall type.

A number of microscopy methods have been used to determine the magnetic texture, domain wall type and to characterise length scales of both domain walls and skyrmions in such multilayer systems. These include magnetic force microscopy (MFM) [20,21] spin polarised scanning tunnelling microscopy (SP-STM) [22], scanning transmission X-ray microscopy (STXM) [10,23], spin polarised low energy electron microscopy (SPLEEM) [24], magnetic transmission soft X-ray microscopy (MTXM) [23] and X-ray photoemission electron microscopy (X-PEEM) [25]. The technique of Lorentz TEM is used in this study and has previously been utilised in a number of investigations into the structure and magnetic behaviour of domain walls and skyrmions in chiral systems [26-28]. Specifically, it has been used to identify whether walls in multilayer materials with chiral texture are of Néel type [29] and also to spatially resolve domain wall widths which may only be ten nanometres or less [30]. CD-XRMS, X-PEEM, SPLEEM and (SP)STM are all surface-sensitive techniques revealing magnetization texture in the top layer(s) only. Lorentz TEM is advantageous as it is sensitive to all layers, providing an averaged projection of the magnetic textures through the thickness. Combining information from these different techniques allows one to build a three-dimensional model of the magnetization texture. In this paper we detail quantitative measurements of the integrated induction using the methods of Lorentz TEM, confirming that the walls in the multi-repeat system studied here are truly hybrid walls with both Bloch and Néel components.

Results

Three multilayer samples were prepared by dc magnetron sputtering. The layer structure studied is Ta(10)|Pt(8)[Co($T$)|Ru(1.4)|Pt(0.6)]×$N$|Pt(2.4), where the numbers are the layer thicknesses in nm and the substrate is on the left hand side. The parameters $T$ and $N$ are the magnetic thickness in nm and the number of repeat layers respectively (with combinations ($T;N$) = (1.2;5), (1.4;10) and (1.6;15)). In the following, for convenience, we refer to the three samples by the number of magnetic layers present only, *i.e.* 5×, 10× and 15×. Note however that reported properties do depend on the details of the structure, and not only $N$. Alternating gradient field magnetometry (AGFM) suggests that all three samples support out-of-plane domains; for the 5× and 10× samples the origin is PMA from the interface but in the 15× sample, because of the Co thickness, the magnetic anisotropy favours in-plane magnetization and out-of-plane domains are stabilised by dipolar interactions. The 1.4 nm Ru layer is used in conjunction with varying Co layer thickness to ensure ferromagnetic (RKKY) coupling between the individual Co layers in each multilayer [31]. As RKKY coupling is an interfacial effect the samples with thinner Co layers experience the strongest interlayer coupling: the 15× sample (with 1.6 nm Co layers) has a weaker coupling than the 5× sample (with 1.2 nm Co layers). However, in any case, the interlayer exchange coupling is small compared to the intra-layer direct exchange $A$. For the TEM studies the samples were deposited on thin $Si_3N_4$ membranes which formed electron transparent 100 μm square windows on a thicker opaque silicon support. From previous work [17,32] and as will be detailed later, it is expected that the 5× is likely to support only Néel type walls whereas the 10× and 15× samples are expected to have hybrid domain walls, i.e. with both Néel and Bloch contributions.

Before presenting the experimental Lorentz TEM images, we first set out the expected difference between imaging pure Néel and hybrid Néel-Bloch domain walls. We also detail the reasons that the quantitative analysis confirms the contribution from the Bloch and Néel components in each case.

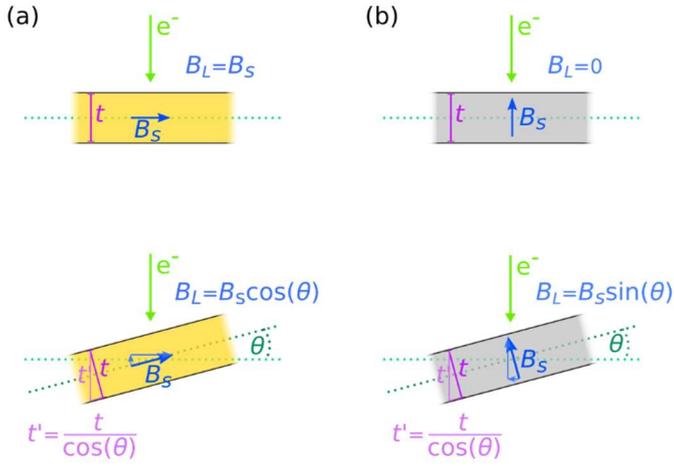

Figure 1. (a-b) Schematic figures to show to orientation of the electron beam in the TEM for films with (a) planar and (b) perpendicular magnetisation. In both cases the films have saturation induction, $B_s$, and thickness, $t$, and are shown untilted and tilted by an angle $\theta$. The projected thickness is indicated as $t'$. Lorentz microscopy is sensitive to the magnetic induction perpendicular to the trajectory of the electron beam, this quantity is labelled as $B_L$.

Firstly, we discuss the Fresnel mode of Lorentz TEM whereby the imaging lens is defocused to reveal domain walls as black and white lines, visible due to the Lorentz deflection of the electron beam. For materials with out-of-plane domains this often requires tilting the sample from its normal position (where the film plane is perpendicular to the electron beam). Tilting results in a deflection of the beam due to the Lorentz force in the out-of-plane domains, and therefore provides contrast. This has been considered previously but only when walls have been of pure Néel or Bloch type [29]. In the case of hybrid walls, we need to consider how both components contribute to the Lorentz deflection. This is done with the aid of Fig. 1(a) and (b) which illustrate the effect of tilting on the two components of induction. Assuming magnetic films with constant saturation induction in each case, the important consideration in Lorentz TEM is the induction component perpendicular to the electron beam ($B_L$ in the figures) integrated along its path. Therefore, for a sample with in-plane magnetisation of thickness $t$ and saturation induction $B_s$, we see that at normal incidence this integrated induction will be $B_st$ - as shown in upper Fig. 1(a). However, when the sample is tilted by an angle $\theta$, as illustrated in lower Fig. 1(a), the orthogonal induction component becomes $B_s\cos\theta$. In this case, the projected thickness presented to the beam is $t/\cos\theta$ resulting in a total integrated induction which is also $B_st$. By contrast, an untilted film with out-of-plane domains gives an integrated induction component which is zero (Fig 1(b) upper) as the induction vector is parallel to the electron beam. When tilted (Fig. 1(b) lower) the component of induction and the projected thickness become $B_s\sin\theta$ and $t/\cos\theta$ respectively, meaning the integrated induction is then $B_st\tan\theta$. With this information we now discuss what to expect from calculated Fresnel images.

In order to do this, we present a model of the magnetisation of a closely spaced pair of domain walls (three domains) which is shown in Fig. 2(a-c), we chose this configuration as it matches the best conditions for the experimental imaging of worm domains under external out-of-plane magnetic field. The walls themselves are separated by 60 nm and each wall was created with a 1D $\tanh(x/\Delta)$ function with a width parameter of $\Delta = 15$ nm, a typical value. The model for pure Néel type walls is shown in the top half of Fig. 2(a-c). The model for a hybrid wall with 10% Bloch to 90% Néel (representing, for example, a ten-repeat multilayer with nine layers Néel type and one layer Bloch type) is shown in the lower half of Fig. 2(a-c) and has $|M_x| \leq 0.9M_s$ and $|M_y| \leq 0.1M_s$. These are simple one-layer, thickness-averaged models as in Lorentz TEM we measure a projection of magnetic induction through the thickness. Thus, it must be noted that a Fresnel image of a hybrid domain wall (varying $M_x$ and $M_y$ through the thickness) is identical to an intermediate domain wall (constant but non-zero $M_x$ and $M_y$ through the thickness) – sketches of these two configurations are provided in supplementary information S1. The calculated Fresnel images for Néel and hybrid domain walls are shown in Fig. 2(d-f) for both the untilted sample and also where the sample is tilted. Here the tilt axis is perpendicular to the length of the walls and in the plane of the film as indicated.

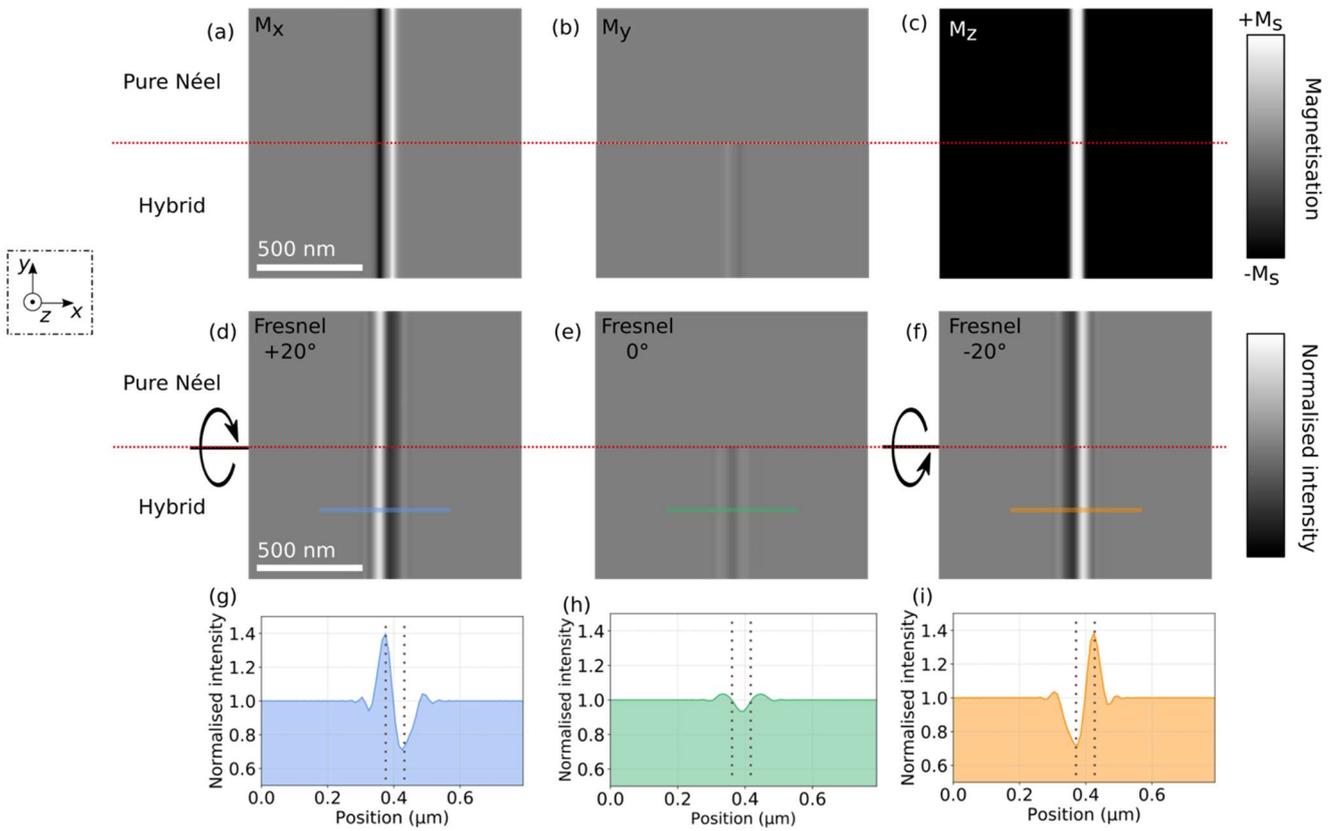

Figure 2. (a-c) $M_x$ $M_y$ and $M_z$ components, constructed from a simple 1D hyperbolic tangent model, of two closely spaced domain walls. The upper half, above the red dividing line, (a-c) models pure Néel walls whilst the lower half models hybrid Bloch/Néel walls, with a Bloch to Néel ratio of 0.1 to 0.9. Calculated Fresnel images (d-f) of the pure Néel (upper) and hybrid (lower) wall taken at tilt angles of +20°, 0° and -20° respectively, about axis indicated by arrowhead. Intensity line traces from the lower part of images (d-f) are shown in (g-i). The dashed lines on the line traces correspond to the centre positions of the domain walls.

It has already been proven that the $M_x$ component of a pure Néel wall gives no image contrast because the wall magnetisation is divergent [29]. The magnetisation of such a wall has an associated demagnetizing field which results in net zero integrated induction – thus its only contribution is an effective reduction of $B_s$. Therefore, as shown in the upper part of Fig. 2(d-f), only contrast arising from the $M_z$ component is observed when the sample is tilted. At zero tilt, Fig. 2(e), no contrast is observed from the pure Néel walls. Additionally, contrast is seen to reverse when tilted in opposite directions, compare upper Fig. 2(d) and (f). In the case of the hybrid walls, the calculated contrast is shown in the lower part of Fig. 2(d-f). With simple visual inspection the images of tilted samples with pure Néel and hybrid walls, Fig. 2(d,f), appear indistinguishable. However, line profiles indicate that peaks of left and right domains have different magnitude due to the Bloch contribution. There is an observable difference between the untilted images, Fig. 2(e), with contrast visible in the lower part of Fig. 2(e) from the hybrid walls only, although this contrast is notably weaker than in the tilted images. These differences and similarities between Fresnel images of pure and hybrid walls are explained by considering how the integrated magnetic induction of each component behaves with tilt – as was shown and discussed regarding Fig. 1. In the case of the Néel wall, the out-of-plane component at a tilt of 20° will give an effective integrated induction of $0.36B_st$. Whereas in the case of the hybrid wall at zero tilt, the contrast arises from the Bloch component ($0.1M_y$) which translates to an integrated induction of $0.1B_st$. This contribution will not change with the tilt and so the out-of-plane component dominates the contrast at the tilts shown, explaining why the tilted images of pure Néel walls and hybrid walls appear indistinguishable by eye. The signature contrast in the untilted hybrid case arises entirely from the Bloch component of the wall. The magnitude of the contrast is important in helping to identify that it is a hybrid wall and not a purely Bloch wall

– if it was purely a Bloch wall the contrast would be ten times higher than seen in Fig. 2(e) (i.e. would be due to $B_s t$ rather than $0.1 B_s t$).

Experimental images were obtained in Fresnel mode using a JEOL ARMcF operated at 200kV [33]. Additionally, we used a pixelated detector, Medipix3, to acquire images rather than a traditional CCD camera due to superior noise performance [34]. In Fig. 3, we show a low magnification image from the 15× sample with an applied out-of-plane field of 270 mT in an untilted orientation with respect to the opaque silicon frame that forms part of the substrate; part of the frame is visible in this image as the large black areas. Applying this field results in long isolated worm domains corresponding to wall pairs which are good for contrast analysis, much more convenient than the demagnetised state for which domain walls are present densely all over the sample.

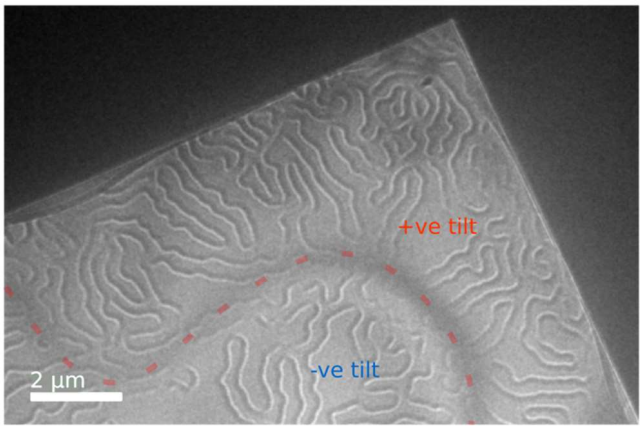

Figure 3. Low magnification Fresnel image of 15× layer sample with closely spaced pairs of domain walls. The sample is in the untilted position however the film on the transparent membrane is buckled and therefore tilted with respect to the beam. The dashed red line marks a bend contour and shows where the effective tilt with respect to the electron beam changes from negative to positive and the contrast is seen to reverse.

Interestingly, although the sample is nominally untilted, we notice that magnetic contrast is clearly visible throughout the membrane. It looks to be from narrow reverse domains which have been nucleated in a worm-like geometry. This indeed is very similar to the wall model shown in Fig. 2. However, it is important to note that distinct black/white contrast that can be seen in Fig. 3 is reversed in two regions separated by a diffuse dark line which is indicated by a dashed red line in the image. This s-shaped boundary is a bend contour and its presence indicates that the flexible membrane shows buckling in this region, which effectively results in a local tilting of the film with respect to the electron beam with the red line separating areas of opposite local tilt. This conclusion is supported by atomic force microscope (AFM) data given in supplementary information S2 which confirms membrane surface tilts approximately ± 15° relative to the silicon frame. Thus, the two areas either side of the red line in Fig. 3(a) show contrast of the same nature as observed in the tilted images of the model in Figs. 2(a) and (c) and Figs. 2(d) and (f). This shows that the contrast here is dominated by the out-of-plane domains, as expected for either Néel or hybrid walls with a small Bloch contribution.

By using a tilt-rotate specimen rod in the TEM we can orientate the sample in a direction that allows us to vary the tilt perpendicular to the wall length and observe how the contrast changes. The absolute value of the tilt of the multi-layered structure with respect to the beam thus corresponds to the local tilt due to buckling, added to the microscope rod tilt. By slowly varying the tilt, we were able get a reversal of the contrast from the out-of-plane domains and then image the area at the cross-over point, which was taken as the effective untilted image. The Fresnel images at effective positive, zero and negative tilt are shown in Figs. 4(a-c). Note that the "zero" tilt image was at a tilt of -9.8° with respect to the notional flat plane of the membrane (i.e. the silicon opaque surface). So only the area indicated by the red oval is at zero tilt with the rest of the field of view at various levels of positive tilt with respect to the beam, this is again due to the buckling of the membrane. In Fig. 4, a clear reversal of the contrast is visible, white/black in Fig. 4(a) and black/white in Fig. 4(c). In Fig. 4(b) the contrast in the area indicated by the red oval is indeed going through a change along the wall pair. Moreover, the intensity here is considerably less than for the tilted images but there is a clear signal from the magnetic structure present.

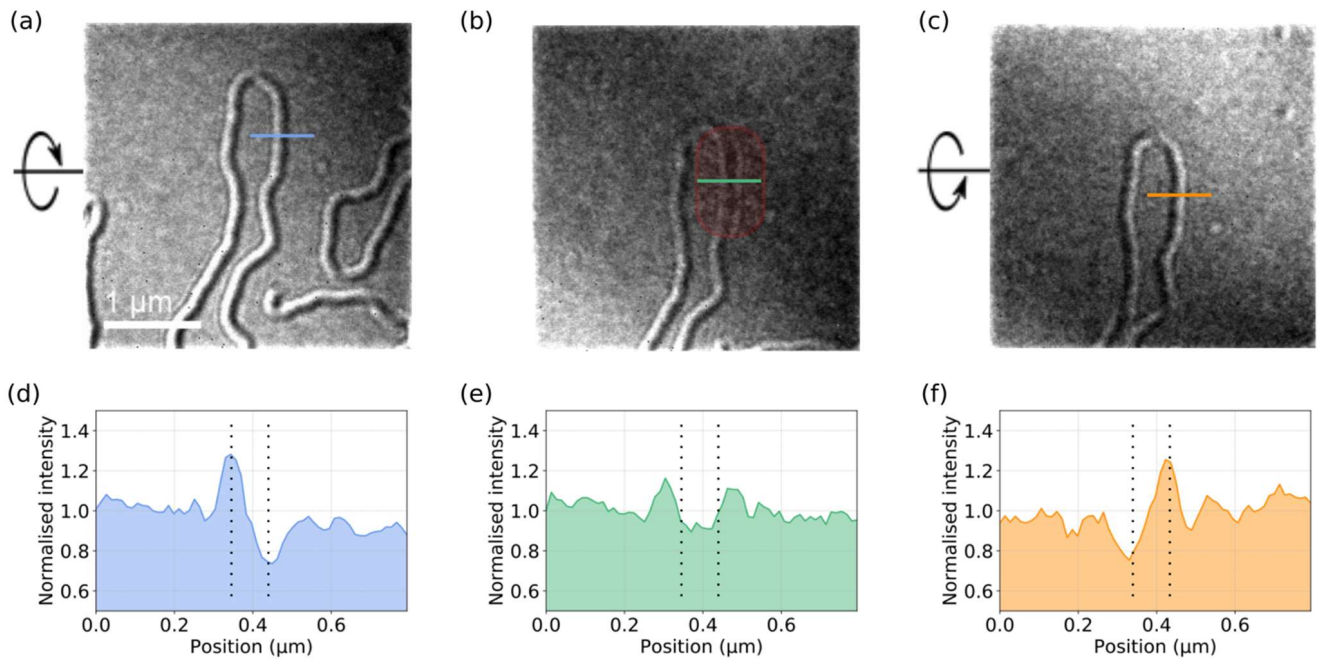

Figure 4. Fresnel images of 15× layer sample with closely spaced pairs of domain walls with varying local tilt at the coloured lines to be (a) +5°, (b) 0° and (c) -5°. The corresponding intensity line traces from these images, averaged over 10 lines, are shown in (d-f). The dashed lines on the line trace are to guide the eye and indicate the central position of the domain wall with respect to the Fresnel image contrast. In Fresnel images the important feature is the contrast level relative to the background, therefore the line traces have been normalised to the background value.

To illustrate the variation, intensity line traces taken from these three images are shown in Figs. 4(d-f) with the contrast normalised to the background intensity to allow for direct comparison of the wall contrast in the three images. These line traces are in excellent agreement with the model suggested for the hybrid wall structure in Fig. 2. In themselves these Fresnel images are not quantitative and do not unambiguously prove that the walls have a hybrid structure. However, they do indicate that there is Bloch character to the walls and they are not of purely Néel type. The reduced intensity certainly suggests that the walls may be hybrid, but we rely on quantitative Lorentz imaging and micromagnetic simulations, described in the next section, to prove this. Fresnel images were also taken of the 10× and 5× samples; these are shown and discussed in supplementary information S3&4. In summary, the images of the 10× sample appear similar to the 15×, showing a definite degree of Bloch character, where the images from the 5× sample appear consistent with a pure Néel wall structure.

To quantitatively determine the Bloch component of hybrid domain walls, the samples were imaged using the technique of differential phase contrast (DPC) in scanning TEM (STEM), this also utilised the Medipix3 detector [35]. Using this detector as opposed to a standard quadrant detector allows a more precise measurement of the shifts of the unscattered central diffraction disk due to the Lorentz deflection of the electron beam. In the case of multilayer films with out-of-plane domains, experience shows that the pixelated detector is necessary to get good magnetic contrast, especially for polycrystalline films with small Lorentz deflection angles [30]; the difficulties related to imaging the multilayer samples specific to this study are discussed in supplementary information S5. The beam shifts measured by DPC are converted into quantitative integrated induction maps which allow quantification of the contributions from the out-of-plane domains and Bloch wall components.

In order to use DPC to determine quantitatively the Bloch contribution to the hybrid domain walls, two image sets were acquired. One at local zero tilt and the other of the same area but obtained at a non-zero relative tilt to the first image. For hybrid walls there are two possible contributions to the Lorentz deflection and hence integrated induction as detailed in Fig. 1. Firstly, there is the contribution from the Bloch walls, $B_s t_B$, where $t_B$ is the film thickness in the system associated with the Bloch wall structure. Secondly, there is the tilt dependent contribution from the out-of-plane domains, $B_s t \tan\theta$, where $t$ is the total magnetic thickness. Assuming the tilt perfectly corresponds to normal electron incidence, the first image contains only a contribution from the Bloch walls. The second image is taken at a tilt angle $\theta$ relative to the first image, where $\theta$ is chosen to be sufficiently large so that the domain contrast clearly dominates. This second image is then used as a reference from which $B_s t$ is extracted.

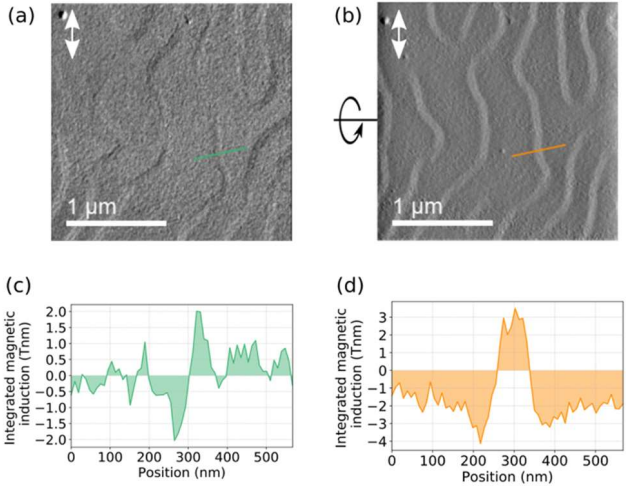

Figure 5. DPC images of the 15× layer sample taken at two different values of sample tilt. The local tilt at the coloured lines are (a) 0° and (b) +13.2°. The component of induction mapped is indicated by the double headed arrow. The corresponding line traces, averaged over 20 lines, of the integrated induction from these images are shown in (c) and (d).

DPC images from the 15× sample are shown in Fig. 5 at two different tilts and map the component of integrated induction in the direction indicated by the double-headed arrow inset on Figs. 5(a-b) which is close to the orientation of the wall length. Again, the zero tilt was found in the Fresnel mode prior to DPC imaging and the region at the green line in Fig. 5(a) was determined as untilted (the tilt with respect to the flat silicon substrate was 12.2°). The image in Fig. 5(b) shows the sample tilted a further 13.2° with respect to Fig. 5(a). Line traces showing the integrated induction from the same regions are given in Figs. 5(c) and (d). In Fig. 5(a) and the associated line trace in Fig. 5(c) the contrast shows a black/white contrast at the walls with the intensity change corresponding to an integrated induction $\pm B_s t_B$. For the tilted image shown in Fig. 5(b) and the associated line trace in Fig. 5(d) the integrated induction corresponds to $\pm B_s t \tan\theta$. As $B_s$ is the same throughout the film and knowing the value of the tilt, $\theta$, the integrated induction ratio from these two images provides the ratio $t_B/t$ to be 0.18 ± 0.02. Converting this into magnetic layers suggests an equivalent of 2.7 of the 15 layers constitute the Bloch portion of the hybrid wall. The measurements from the 10× sample give a very similar ratio of 0.16 ± 0.02 and equates to 1.6 of the 10 layers having Bloch nature. As was previously mentioned, the methods of Lorentz microscopy do not distinguish between an intermediate Néel/Bloch wall (one with non-zero but constant $M_x$ and $M_y$ throughout the thickness) and a hybrid Néel/Bloch wall (one with varying $M_x$ and $M_y$ through the thickness) as the thickness projected magnetic induction from each is identical. However, taken in conjunction with the surface sensitive measurements by x-ray magnetic scattering experiments from similar samples in [17] - which measured opposite handedness of Néel walls on, effectively, the top and bottom surfaces of the samples – this allow us to be certain we are imaging hybrid domain walls. As with the Fresnel images we were unable to obtain any measureable Bloch signal from the 5× sample at zero tilt, therefore this suggests a Bloch contribution below our measurement capability or is consistent with the wall being purely Néel in character. For comparison, we performed micromagnetic simulations with parameters corresponding to these three samples - details are provided in the methods section. Projecting the magnetisation through the thickness, the ratio $t_B/t$ was found to be 0.04, 0.16 and 0.19 in the simulations of the 5×, 10× and 15× samples respectively. This is in excellent agreement with the experimentally measured values of 0.16 and 0.19 found for the 10× and 15× samples and is also consistent with the lack of evidence of hybrid walls in the 5× sample. The small portion of Bloch character predicted for the 5× sample corresponds to a very slight twist away from a pure Néel configuration in the top most layer of the simulation and is not an indication of a hybrid type wall. This twist appears in simulations where the DMI energy is only just stronger than the dipolar effects and is likely below our measurement capability. The cross-sectional profiles of the walls from micromagnetic simulations are presented in supplementary information S6.

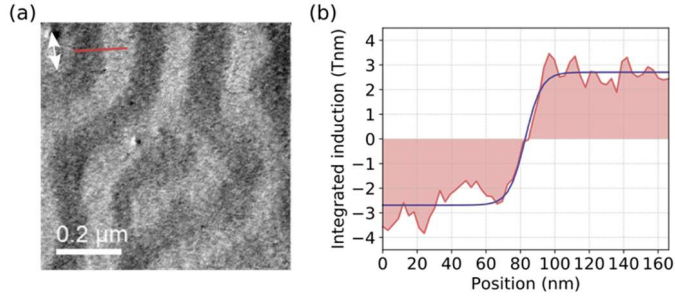

Figure 6. (a) DPC image taken of the 15× sample tilted and in a demagnetized state close to remanence, with component of indication mapped indicated. The sample is tilted to provide strong contrast from the out-of-plane domains. (b) Integrated induction line trace taken from red line in (a) showing domain wall profile and its fit to a hyperbolic tangent function, Δ, see text for details.

The spatial resolution of DPC allows the profile of the narrow domain walls to be imaged and the wall width to be measured. In Fig. 6(a), we show a DPC image of the 15× sample in a demagnetised state close to remanence together with an integrated induction profile in Fig. 6(b). After deconvolution with a Gaussian function matching the imaging probe, the line trace can be fitted to a standard hyperbolic tangent function, $B_y = B_s \tanh(x/\Delta)$, giving a measure of the thickness-averaged domain wall width by the fit parameter Δ. This procedure identified an average $\Delta = 11 \pm 1$ nm in the 15× sample and $\Delta = 5 \pm 1$ nm in the 10× sample. The DPC images from the 5× sample contain strong polycrystalline contrast which obscures the magnetic contrast and prevents reliable extraction of the wall width – this is due to the small magnetic thickness and the issues discussed in supplementary information S5.

| Sample | Experiment | | Simulation | |
|---|---|---|---|---|
| | Δ (nm) | $t_B/t$ | Δ (nm) | $t_B/t$ |
| 5× | - | - | 4 | 0.04 |
| 10× | 5 ± 1 | 0.16 ± 0.02 | 5 | 0.16 |
| 15× | 11 ± 1 | 0.18 ± 0.02 | 10 | 0.19 |

Table 1. Table of the width parameter, Δ, and ratio of Bloch thickness to total magnetic thickness, $t_B/t$, measured from each sample experimentally and calculated from micromagnetic simulations of each sample – details of the simulations are provided in the methods section.

The salient experimental results - both the wall width parameter, Δ, and the Bloch thickness to total magnetic thickness ratio, $t_B/t$ - are summarised in Table 1, along with the same quantities measured from micromagnetic simulations. The micromagnetic simulations are in excellent agreement with the experimental measurements from both the 10× and 15× samples whilst, as mentioned previously, the wall width cannot be measured from the 5× sample but the very small $t_B/t$ ratio is consistent with the lack of magnetic contrast at observed at zero tilt in experimental images. The details of the micromagnetic simulations are provided in the methods section. Experimental determination of the domain wall width is of particular interest as it allows indirect access to the exchange stiffness, $A$, of the material that is the only material parameter not experimentally measured for input to micromagnetic simulations. In supplementary information S7 we explore, with micromagnetics, how Δ evolves as $A$ is varied between 3 and 20 pJm$^{-1}$. Fitting the experimentally measured Δ values to the curves of Δ and $A$ suggests, for both the 10× and 15× samples, $A$ in this material is close to 12 pJm$^{-1}$.

Discussion

Here we present a Lorentz microscopy study which allowed an unambiguous and quantitative determination of the nature of hybrid domain walls in ultrathin multilayer stacks. These measurements are consistent with recent studies which have measured the Néel component and also with the simulations of such structures [17-19]. Our investigations also provide direct measurement of the distinct Néel and Bloch contributions to these hybrid walls. As noted in previous Lorentz TEM studies of ultrathin films with DMI, imaging of the Néel components is somewhat indirect, however we have been able to measure directly the Bloch contribution, aided in part by simple models and micromagnetic simulations of the domain wall structures. Our findings show that only a small fraction, about one fifth, of the layers in the system studied have a Bloch configuration when the number of layers $N = 10$ and 15. When the number of layers is small ($N = 5$), this Bloch rotation appears not to be present. Note that the critical number of repetitions to generate hybrid textures depends not only on the number of repetitions but also upon other magnetic parameters such as effective magnetization, magnitude of D, etc. For all three multilayer systems studied, our experimental results are in excellent agreement with micromagnetic simulations for both the ratio of Bloch to Néel wall contributions and the domain wall widths. The predicted width parameters are 10 nm and below, values that can easily be measured by the high spatial resolution Lorentz microscopy techniques. As an aside we note that the observed magnetisation configuration seen through the thickness in these multilayer films resembles the structure in walls observed previously in thick (> 60 nm) single layer magnetic films almost 50 years ago but possessing planar magnetisation [36,37]. Here however the multilayer structure, with limited stiffness along the out-of-plane direction, allows twisting of the DW texture even for relatively modest thickness. Additionally, thickness dependent wall structures have also been predicted and observed in other studies of skyrmions in bulk materials. Two examples are so-called chiral bobber structures as well as Néel caps in bulk Bloch skyrmions [38-41] – it should be noted that the neither are driven by dipolar factors. Detailed structure of the domain walls is extremely important for stability and dynamics of the domains. We demonstrated here that the metrology provided by nanoscale quantitative imaging is necessary to predict how spin polarized currents will interact with chiral domain walls for spintronic applications [17,18].

Methods

Sample preparation
Multilayers have been deposited by dc magnetron sputtering at room-temperature, under Ar gas flow at a pressure of 0.25 Pa. The deposition rates were calibrated prior to the present depositions by X-ray reflectivity measurements. Base pressure of the sputtering system was better than $8 \cdot 10^{-6}$ Pa. The multilayers of this study have been deposited on top of Ta(10 nm)/Pt(8 nm) buffers, which allows a control over their perpendicular magnetic anisotropy, and capped with Pt(3 nm) layers to prevent oxidation. The saturation magnetisation $M_s$ = 1-1.2 MA/m has been obtained by averaging SQUID measurements.

Lorentz and atomic force microscopy
All of the Lorentz microscopy images were taken using a JEOL ARM 200cF equipped with a cold field emission gun and CEOS probe aberration corrector. The Fresnel imaging was carried out in TEM mode with lens defocus of between 1 and 5 mm was used. The DPC imaging was performed in STEM with a condenser (probe forming) aperture of diameter 40 μm. The latter gives a probe of 3.5 nm and a resolution of 1.75 nm. The images providing the high-resolution domain wall profiles were acquired with a sampling pixel size of 3.0 nm at the highest magnification used here of ×250k. The AFM data in the supplementary information was taken using a Veeco Dimension 3100 Scanning Probe Microscope operated in tapping mode with a standard non-magnetic tip. The AFM image displayed in this paper is a scan over an area of 30×30 μm.

Micromagnetic simulations
Domain profile simulations were performed with Mumax3 [42]. The cell size was fixed as 0.25x0.25x0.2 nm along x (wall normal), y (wall length) and z (layer planes normal) directions, respectively. For demagnetized systems, a unique period of domains was simulated, which was extended through periodic boundary conditions in the x and y directions, extending the system to 32 domain periods and 4.096 μm along x and y directions respectively. Then the magnetic configuration composed of one up domain ($M_z = M_s$) and one down domain ($M_z = -M_s$) separated by domain walls (using the domain periodicity measured by Lorentz TEM) was directly relaxed to reach the minimum energy state. Parameters were Heisenberg exchange $A = 10$ pJ/m and saturation magnetisation $M_s = 1$ MAm$^{-1}$ for the 5× and 10× samples and $M_s = 1.2$ MAm$^{-1}$ for the 15× sample. Simulations were trialled with different values of $M_s$, over the range identified experimentally by SQUID, the above values were decided on as they provided the best fit to the experimental data. The magnetic thickness of the $N$ layers and thickness of vacuum spacing are chosen to match the ones of each multilayer. Uniaxial perpendicular anisotropy $K_u = 0.829$ MJm$^{-3}$, 0.711 MJm$^{-3}$ and 0.622 MJm$^{-3}$; DMI parameter $D = 0.825$ mJm$^{-2}$, 0.707 mJm$^{-2}$ and 0.619 mJm$^{-2}$ were chosen for samples with Co layer thickness $T = 1.2$ nm, 1.4 nm and 1.6 nm respectively. They thus match with the interfacial perpendicular anisotropy and interfacial DMI value measured from a similar sample with $T = 1.1$ nm considering an inverse thickness dependence.

Additional Information
The authors declare no competing financial and non-financial interests.
Raw data used to provide the results in this report can be found at:
http://dx.doi.org/..................................................

Acknowledgements
The authors acknowledge financial support from European Union grant MAGicSky No. FET-Open-665095 and EPSRC through grant EP/M024423/1 and from the Agence Nationale de la Recherche, France, under Grant Agreement No. ANR-17-CE24-0025 (TOPSKY). Additionally, we thank Albert Fert who was initially involved in devising these film structures.

Author contributions
W.L. V.C. and N.R. conceived the project and the material design. W.L. F.A., and S.C. prepared the films. S.McV. and K.F. planned the TEM/STEM experiments and K.F. acquired and analysed the imaging data. S.McV. and K. F. devised the wall model for TEM image interpretation and calculation. W.L. undertook the micromagnetic simulations. K.F., S.McV., W.L., V.C and N.R. prepared the manuscript. All authors discussed and commented on the manuscript.